\documentclass[12pt]{article}

\usepackage[utf8]{inputenc}
\usepackage{amsmath,amsthm,amssymb}
\usepackage{enumerate}
\usepackage{graphicx}
\usepackage{epsfig}
\usepackage{color}
\usepackage{hyperref}
\usepackage{bm}
\usepackage{caption}
\usepackage{subcaption}
\usepackage{authblk}
\usepackage{float}
\graphicspath{{pics1/}}
\title{{$\mathcal{PT}-$Symmetry and Supersymmetry: Interconnection of Broken and Unbroken Phases}}

\author[1]{\small {Adipta Pal}\thanks{adiptapal7@gmail.com}}
\author[2]{Subhrajit Modak \thanks{ modoksuvrojit@gmail.com}}
\author[3]{Aradhya Shukla \thanks{aradhya02@iiserkol.ac.in}}
\author[3]{Prasanta K. Panigrahi\thanks{pprasanta@iiserkol.ac.in}}

\affil[1] {\small{Max Planck Institute for the Physics of Complex Systems,\vskip 0.1cm N\"othnitzer Stra{\ss}e 38, D-01187 Dresden, Germany }} \vskip 0.1cm
\affil[2]{ Indian Institute of Science Education and Research Mohali,\vskip 0.1cm Punjab - 140306, India}
\affil[3] {\small{Indian Institute of Science Education and Research Kolkata, Mohanpur,\vskip 0.1cm West Bengal - 741246, India }}
 
\date{}
\begin{document}

\maketitle 
\begin{abstract}
The broken and unbroken phases of $\mathcal{PT}$ and supersymmetry in optical systems are explored for a complex refractive index profile in the form of a Scarf potential, under the framework of supersymmetric quantum mechanics. The transition from unbroken to the broken phases of $\mathcal{PT}-$symmetry, with the merger of eigenfunctions near the exceptional point is found to arise from two distinct realizations of the potential, originating from the underlying supersymmetry. Interestingly, in $\mathcal{PT}$-symmetric phase, spontaneous  breaking of supersymmetry occurs in a parametric domain, possessing non-trivial shape invariances, under reparametrization to yield the corresponding energy spectra. One also observes a parametric bifurcation behaviour in this domain.  Unlike the real Scarf potential, in $\mathcal{PT}-$symmetric phase, a connection between complex isospectral superpotentials and modified KdV equation occurs, only with certain restrictive parametric conditions. In the broken $\mathcal{PT}$-symmetry phase, supersymmetry is found to be intact  in the entire parameter domain yielding the complex energy spectra, with zero-width resonance occurring at integral values of a potential parameter.
\end{abstract}

\section{Introduction} 
The search for new materials with desired optical properties 
has been one of the main driving forces of research in optics and other related areas in recent times \cite{ye, for, park}. Emergence of Parity-Time ($\mathcal{PT}$)-symmetric \cite{BB98,BBJ}  optical structures  represent a new generation of artificial optical systems \cite{GMC, GSD}, which utilize gain and loss in a balanced manner to achieve  desired order of functionality \cite{PTS, parto}. Among these developments, $\mathcal{PT}-$symmetric laser \cite{laser}, the realization of anti $\mathcal{PT}$-symmetry in microcavity \cite{PRL, gsa3} and the ultrasensitive electromagnetic sensors \cite{sensors}  are worth mentioning. The  careful balancing of loss and  gain is now being overcome  through  gain-free route by using  appropriate designed passive systems \cite{laks, virtual}.  Recent observation of unbroken and broken phases of $\mathcal{PT}-$symmetry in photonic condensates near equilibrium condition, has opened further avenues for both physical and technological applications \cite{fahri}.  Interestingly, supersymmetry, a proposed symmetry interconnecting fundamental bosons and fermions \cite{wess}, has been successfully used for controlled lasing \cite{heinri, macho, hok} and other optical applications \cite{RMG1}. Supersymmetry has been found very useful in phase-matched mode selection, removal of undesired modes and  other phase-sensitive applications \cite{miri}. It plays a crucial role in designing $\mathcal{PT}$ inspired optical couplers \cite{principe, laks1}. Both these symmetries are characterized by unbroken and broken phases, having different physical properties in optical systems. Balancing of gain and loss plays the key role in realizing the broken and unbroken phases of $\mathcal{PT}-$symmetry, dramatically changing behaviour of light propagation  in the medium \cite{mak, kep}. At the symmetry breaking point, light can be brought to a complete stop \cite{gold}. Supersymmetry in  optical and quantum mechanical systems, interconnects even and odd eigenfunctions of two distinct systems \cite{khare3, meca}. Interestingly, an additional symmetry, known as shape invariance (SI) characterizes certain systems, where the potential or the refractive index profiles of the supersymmetric partners have same functional form, differing only in their constant parameters \cite{ged, asim3}. The energy values are analytically calculable in this case, allowing one to tune the potential parameters to achieve a desired level \cite{SQM}. The potentials can be further tuned using  the iso-spectral deformation, which enables one to modify the wavefuntion and the potential profile, keeping the eigenvelues intact. This has found use in explaining the resonance width in physical systems \cite{dhruba} and tailoring of the optical fibers for desired beam profile \cite{mak1, alex, pkp1}. Expectedly, a system with both $\mathcal{PT}$ and supersymmetry can provide significant control of medium properties. As the broken and unbroken phases of these two symmetries depend on the potential parameters, a careful identification of the parameter domain for the two physically distinct  phases is worth investigating.
$\mathcal{PT}-$symmetry came into play after the seminal work  of  Bender  and  Boettcher \cite{BB98, BBJ},  in  which  they suggested  an extension of  quantum  mechanics, where the notion  of Hermiticity of the underlying Hamiltonian is replaced by $\mathcal{PT}$ invariance. As an example, a $\mathcal{PT}-$symmetric Hamiltonian contains a one-dimensional Schr\"odinger operator with a complex potential satisfying $V(x)=V^{*}(-x)$, i.e.,  the  real  and  imaginary  parts  of  the potential must  be symmetric and anti-symmetric functions, respectively. Such non-Hermitian systems exhibit many interesting properties, which are otherwise unattainable in conventional Hermitian systems.  One of the most striking $\mathcal{PT}$ properties is the  appearance  of  a  sharp,  symmetry-breaking  transition \cite{AP},  once  the  non-Hermitian  parameter crosses a certain threshold. This transition signifies the occurrence of a spontaneous $\mathcal{PT}-$symmetry breaking from the exact phase, where all of the eigenvalues are real, to a broken phase where eigenvalues  occur  in  complex-conjugate pairs. This symmetry breaking point is termed as exceptional point, where  the eigenvalues coalesce and the corresponding eigenvectors collapse on each other.  Exceptional points have been observed in various photonic systems \cite{miri3}, optical waveguides \cite{RMG1} and time domain lattices \cite{rosen}.

 In optical systems, Helmholtz equation replaces the Schr\"odinger equation, enabling the well-established quantum mechanical tools for use in optical systems.  In order to realize complex $\mathcal{PT}-$symmetric structures, this formal equivalence of the quantum mechanical Schr\"odinger equation to the optical wave equation has been exploited by combining symmetric index guiding with an antisymmetric gain/loss profile  \cite{klai, jiri}. The prospect of utilizing both optical gain and loss has emerged as a new paradigm in shaping the flow of light. These include effects like unidirectional invisibility \cite{LFW}, loss-induced transparency \cite{GSD}, band merging \cite{MRP}, optically induced atomic lattices \cite{ZZS}  and improved lasing  as mentioned earlier. The exceptional points (EPs) have  emerged as a new design tool for engineering the response of optical systems, as near these points,  light propagation has a strong parameter sensitivity, which is further affected by subtle changes in the initial condition \cite{nori2}. Such behaviour is characteristic of classical and quantum catastrophes  \cite{SL}. Recent studies have shown that controllable confinement of light can be made possible at EPs \cite{HR}. The physical phenomena near EPs  reveal many interesting features, like negative refractive index \cite{fl} and Bloch oscillation in complex crystals \cite{longhi1, graefe}. For an account of the recent status of $\mathcal{PT}$-symmetry, the interested readers are referred to \cite{CMB}.

Here, we investigate the Helmholtz equation for light  propagation through a medium, having complexified dielectric distribution, obeying $\mathcal{PT}-$symmetry. For explicitness, we consider complex Scarf profile as refractive index, where all the aforementioned features of $\mathcal{PT}-$symmetric system are analytically tractable. The refractive index profile, in the form of a Scarf potential in the Helmholtz equation, has been recently designed  for controlling reflection of light, which at certain parameter value can be completely reflectionless \cite{GA, gopal}. Interestingly, the complex Scarf potential is also characterized by supersymmetry (SUSY), which provides a new avenue for designing partner index profiles with desired characters \cite{jak}. It can serve as a testing platform, where ramifications of SUSY and $\mathcal{PT}$-symmetry can be explored and  utilized  to  realize new class  of  functional structures.

The paper is organized as follows. Section 2  establishes the  connection between the refractive index profile of the Helmholtz equation and $\mathcal{PT}-$symmetric potential in quantum mechanics, together with a brief discussion about fundamentals of supersymmetric quantum mechanics (SUSY QM). In section 3, the complex $\mathcal{PT}-$symmetric potential is introduced and analysed in terms of SUSY methods, while establishing their connection with optics. The phases of  $\mathcal{PT}-$symmetry and SUSY and their interconnection have been discussed in section 4. The spectral structures and parameter domain for SUSY broken phases are derived for the first time for the complex Scarf II potential.  Subsection 4.1 makes a connection between the spectral problem in unbroken phase and modified KdV equation through isospectral deformation. The subsequent subsection shows another feature of the $\mathcal{PT}-$phase transition, bifurcation phenomenon in the parameter space, which enables transition from one superpotential to another, interconnecting $\mathcal{PT}-$symmetry with SUSY.  Section 5 deals with $\mathcal{PT}-$broken phase, having complex eigenvalues and zero-width resonances, with SUSY intact in the entire parameter domain. Section 6 concludes the paper with directions for future work. 

\section{Refractive index profile and supersymmetric quantum mechanics}
We consider the case of plane monochromatic optical wave  incident at an angle $\theta$, in a medium with inhomogeneous dielectric distribution, $\epsilon(x)=\epsilon_b+\alpha(x)$. The inhomogenity is assumed to vanish at spatial infinity, so that $\epsilon(x)\rightarrow\epsilon_b$ as $\vert x\vert\rightarrow\infty$. The problem can be made effectively one dimensional due to the fact that dependence on $y$ is given by $e^{ik_{y}y}$, where $k_{y}=k_{0}\sqrt{\epsilon_{b}}\text{sin}\theta$. Note that $k_{y}$ is same for each layer due to  continuity of tangential fields, leading to propagation along the $x$ direction. The propagation equation for transverse electric waves is the Helmholtz equation,   
\begin{equation}\label{RE}
\frac{d^2\mathcal{E}}{dx^2}+(k_0^2\epsilon(x)-k_y^2)\mathcal{E}=0,
\end{equation}
that can be connected with the Schr\"odinger equation:
\begin{equation}
E=k_0^{2}\epsilon_{b}\text{cos}^2\theta,
\end{equation}
and
\begin{equation}\label{qq}
V(x)=k_{0}^2\epsilon_b-k_{0}^2\epsilon(x).
\end{equation}
The potential $V(x)$ is said to be reflectionless, if it is transparent to waveforms of arbitrary positive energy. The refractive index, $n(x)$, which is the square root of the dielectric function, is related to quantum mechanical potential $V(x)$ (with $\hbar=2m=1$):
\begin{equation}
n^2(x)=n_{b}^2-\frac{V(x)}{k_{0}^2}.   
\end{equation}
 $\mathcal{PT}$-symmetry requires  the refractive index  to obey $n(x)=n^{*}(-x)$ . The $\mathcal{PT}-$symmetric potentials with the above optical connection have been at the forefront of investigations \cite{malo4}. 

 SUSY QM have played a crucial role in revealing interesting facets of various optical systems \cite{miri6}. SUSY relates two sibling Hamiltonians, which factorize as $H_-=A^\dagger A$ and $H_+=AA^\dagger$, with $H_\pm=-\frac{\partial^2}{\partial x^2}+V_\pm (x)$. A connection can be established with the introduction of the superpotential, $W(x)=-\frac{1}{\psi_0(x)}\frac{\partial\psi_0(x)}{\partial x}$ for the given pair of potentials:
\begin{equation}
V_\pm=W^2(x)\pm\frac{\partial W(x)}{\partial x}.
\end{equation} 
Here, $\psi_0(x)$ is the ground state of $H_-(x)$, with the ladder operators given by $A^\dagger=-\frac{\partial}{\partial x}+W(x)$, $A=\frac{\partial}{\partial x}+W(x)$. One notices that $E_n^+=E_{n+1}^-$ and thus, the spectra of partner potentials differ only in the ground state of $H_- (x)$. 
The eigenvalues and wavefunctions can be explicitly obtained for the shape invariant potentials \cite{SQM}. Partner potentials are said to be shape invariant, if they satisfy relations of the form,
\begin{equation}
V_+(x;p_0)=V_-(x;p_1)+R(p_1),
\end{equation} 
where $p_0$ is a constant parameter for a given SUSY potential pair, $p_1$ is a function of $p_0$ and $R(p_1)$ is a constant function independent of $x$. The spectrum of $H_- (x)$ is obtained as
\begin{equation}
E_0^n=\sum_{k=1}^nR(p_k),
\end{equation} 
and the eigenstates are explicitly given by,
\begin{equation} \psi_n^{-}(x;p_0)\propto A^\dagger(x;p_0)A^\dagger(x;p_1)...A^\dagger(x;p_{n-1})\psi^-_0(x,p_n).
\end{equation} 
The SUSY method can be used to determine the energy spectra in a number of $\mathcal{PT}-$symmetric complex potential \cite{AP}, and therefore consequences of $\mathcal{PT}$-symmetry and its breaking can be analytically probed.  Some of the potentials also exhibit broken SUSY and hence provide an ideal venue for investigating the broken and unbroken phases of both the symmetries and their interconnection.  In the following, for explicitness, we consider the refractive index profile in the form of a Scarf potential, which has found applications for its reflectionless behaviour. In the process,  energy spectra in the broken SUSY domain for the complex potential is analytically obtained, through a combination of shape inavriance and SUSY.

\section{ Broken and unbroken $\mathcal{PT}-$symmetric phases} 
The real Scarf II potential is one of the exactly solvable potentials, having both bound and scattering states \cite{khare}:
\begin{equation}\label{scarf}
V(x) = (-A\,(A+\alpha) + B^2) sech^2\, \alpha x + B (2 A + \alpha) sech \,\alpha x\, tanh\, \alpha x,
\end{equation}
where $A, B$ are real constant parameters. It exhibits many interesting features and has been studied extensively in  SUSY quantum mechanics \cite{levai3}, group theory \cite{levai1, levai2} and other approaches \cite{pkp4}.  It is a non-symmetric extension of the well studied P\"oschl-Teller potential \cite{poschl}:  $V(x)= - A^2 \, sech^2  x$, which has been elegantly related to the angular momentum eigenvalue equation \cite{gursey}: 
\begin{equation}\label{gu}
\Big[ -\frac{1}{sin \theta}\,\frac{\partial}{\partial \theta}\Big(sin \theta \frac{\partial}{\partial \theta} + \frac{m^2}{sin^2 \theta}\Big)\Big]\; u^m_l (\theta) = l(l+1)\,  u^m_l (\theta),
\end{equation}
with the mapping $cos \theta = tanh x$: 
\begin{equation}\label{gu1}
\Big[\frac{\partial^2}{\partial x^2} + \frac{l(l+1)}{sech^2 x} \Big]\, u^m_l (x) = m^2 \,u^m_l (x).
\end{equation}
There are two symmetry operations, $m\rightarrow -m$ and $l\rightarrow -(l+1)$, that leave the potential and eigenvalue invariant, which has non-trivial consequences in our analysis.  SUSY and shape invariance, combined with the above mentioned parametric freedom find elegant use in obtaining the eigenvalue spectra in both broken and unbroken phases of SUSY \cite{diaz}.  The Scarf II potential can be obtained from superpotential $W(x) = A \,tanh \,\alpha x + B\, sech \,\alpha x$. 
The complex $\mathcal{PT}-$symmetric potential can be obtained through a simple parametrization $B\rightarrow iB$ \cite{AJP}: 
\begin{equation}\label{scarf1}
V(x) =  (-A(A+\alpha) - B^2) \,sech^2 \alpha x + i\,B (2 A + \alpha)\, sech \,\alpha x\, tanh\, \alpha x.
\end{equation}
for which the eigenspectra and eigenfunctions have been obtained \cite{ZA12}.
 With the goal of investigating the broken and unbroken phases of  $\mathcal{PT}$ and SUSY, we parametrize the generalised complex superpotentials as \cite{AP},
\begin{equation}\label{oo}
W_{\mathcal{PT}}^{\pm}=\big(A\pm iC^{\mathcal{PT}}\big)\,tanh\,\alpha x+\big(\pm C^{\mathcal{PT}}+iB\big)\,sech\,\alpha x,
\end{equation}
where all the potential parameters are real. The supersymmetric  potential ($V_{-}$) corresponding to $W_{\mathcal{PT}}^{\pm}$ is of the form,
\begin{align}
\begin{split}\label{GV}
    V_-(x) ={}& (A\pm iC^{\mathcal{PT}})^2 -\big[(A\pm iC^{\mathcal{PT}})(A\pm iC^{\mathcal{PT}}+\alpha)-(\pm C^{\mathcal{PT}}
          +iB)^2]sech^2\,\alpha x \\
          &-i\,(\pm iC^{\mathcal{PT}}-B)\,\big[2(A\pm iC^{\mathcal{PT}})
          +\alpha)\big]sech\,\alpha x\,tanh\,\alpha x,
\end{split}
\end{align}
which leads to both real and complex spectra, physically obtainable using SUSY and shape invariance. The potential in Eq. (\ref{GV}) in general may not be $\mathcal{PT}-$symmetric, for which the coefficient of the first and second terms must be real and  purely imaginary, respectively. In the present parametrization, these two conditions lead to a unique relation \cite{AP}:
\begin{equation}\label{CPT}
C^{\mathcal{PT}}\big[ 2(A-B)+\alpha\big]=0.
\end{equation}
 For $C^{\mathcal {PT}} =0, \mathcal{PT}-$symmetry is unbroken and one gets bound states with real spectra. However, SUSY is broken in some parametric domains. Nonzero value of $C^{\mathcal {PT}}$ leads to the  spontaneous $\mathcal{PT}$ broken phase, where the energy eigenvalues are in general complex. Therefore, $C^{\mathcal {PT}}$ behaves like an order parameter, whose non-vanishing value characterizes broken $\mathcal{PT}$ phase. Interestingly, in both unbroken and broken phases of  $\mathcal{PT}$, the potential $V_-(x)$ can be obtained from two distinct superpotentials, due to its underlying symmetry. Interestingly, these two superpotentials generate two disjoint sectors of the Hilbert space in the unbroken case. Here the spectrum is real and  shape invariance leads to translational shift in a suitable parameter by real units \cite{AP}. For a different parametrization, two different superpotentials lead to the same potential for the broken $\mathcal{PT}-$symmetry as well. In this case, shape invariance produces complex parametric shifts.  
  We discuss in great detail about the interconnection between unbroken $\mathcal{PT}-$symmetric and broken SUSY as well as spontaneously broken $\mathcal{PT}$ phases in the following sections.

  \section{Unbroken $\mathcal {PT}$ and broken SUSY phases}
For  $C^{\mathcal{PT}}=0$, the potential becomes $\mathcal{PT}-$symmetric,
\begin{align}
\begin{split}\label{VPT}
    V_{\mathcal{PT}}(x) = A^2 -\big[A(A+\alpha)+B^2\big]sech^2\,\alpha x 
               +iB(2A+\alpha)sech \,\alpha x\,tanh\,\alpha x,
\end{split}
\end{align}
derived from the superpotential,
\begin{equation}\label{W1}
W_1(x) = A~tanh\,\alpha x+iB~sech\,\alpha x.
\end{equation}
This potential is invariant under the transformation $A\leftrightarrow B-\frac{\alpha}{2}$ \cite{AP,BQ02, bpm}, which provides another candidate for the superpotential,
\begin{equation}\label{W2}
W_2(x)=(B-\frac{\alpha}{2})\,tanh\,\alpha x+i(A+\frac{\alpha}{2})\,sech\,\alpha x.
\end{equation}
The above mentioned parametric freedom enables two sets of  $SL(2, \mathcal C)$ algebra, which can be  mapped onto each other, yielding  a unique $\mathcal{PT}-$symmetric potential \cite{bag}. These two superpotentials  lead to two sets of real energy eigenvalues,
\begin{equation}\label{E1}
E^1_n=-(A-n\alpha)^2,
\end{equation} 
and
\begin{equation}\label{E2}
E^2_n=-(B-\frac{\alpha}{2}-n\alpha)^2.
\end{equation}
  valid  if, $A>0$ and $B>\frac{\alpha}{2}$, respectively. For other regions in the parameter space, at least one of the energy eigenvalue set fails to give a normalizable ground state, which points to the case of broken supersymmetry. Interestingly, for the broken SUSY cases, exact energy eigenvalues can be obtained by two steps of shape invariance \cite{GMS,RMG}. One first converts the potential to the SUSY unbroken state and then finds the energy eigenvalues. The regions of broken and unbroken SUSY have been depicted in Fig. 1 . The energy eigenvalues in these regions are obtained as follows:
\begin{figure}[t]
\begin{center}
\includegraphics[scale=1.0]{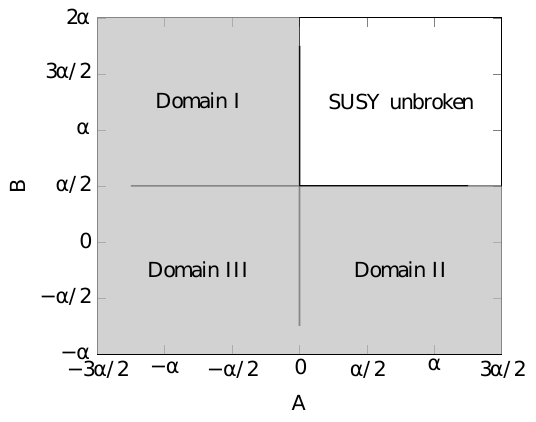}
\caption{The shaded regions I, II and III constitute the broken SUSY areas in the parameter space. The energy eigenvalues for the broken SUSY domains can be obtained by considering two shape invariances of the complex $\mathcal{PT}-$symmetric Scarf II potential.}
\label{fig20}
\end{center}
\end{figure}
\begin{itemize}
\item{\textbf{Domain I} ($A<0,\; B>\alpha/2$): Consider first the shape invariance $A\rightarrow -A$ and then apply the known shape invariance $A\rightarrow A+\alpha$, to get energy eigenvalues for Eq. (\ref{scarf1}) as
\begin{equation}\label{E3}
E^1_{n,I}=-(A+n\alpha)^2,
\end{equation}
$E^2_n$ remains the same. Notice that the first shape invariance changes the sign of $A$ so that unbroken SUSY can be applied.}

\item{\textbf{Domain II}  ($A>0,\; B<\alpha/2$)): Consider first the shape invariance $B-\alpha/2\rightarrow -(B-\alpha/2)$, $A+\alpha/2\rightarrow -(A-\alpha/2)$ and then apply the known shape invariance $B-\alpha/2\rightarrow B-\alpha/2+\alpha$, $A-\alpha/2\rightarrow A-\alpha/2$, to get energy eigenvalues  as,
\begin{equation}\label{E4}
E^2_{n,II}= -(B-\frac{\alpha}{2}+n\alpha)^2.
\end{equation}
$E^1_n$ remains unchanged. The first shape invariance changes the sign and takes it to the unbroken SUSY domain, sign change for the second parameter does not affect as it does not contribute to the energy for Eq. (\ref{W2}). 
}

\item{\textbf{Domain III} ($A<0,\; B<\alpha/2$): Here one needs to consider both the techniques, one for Domain I for the eigenvalues of Eq. (\ref{W1}) and one for Domain II for the eigenvalues of Eq. (\ref{W2}), which yields
\begin{equation}\label{E5}
E^1_{n,III}=-(A+n\alpha)^2,\qquad E^2_{n,III}=-(B-\frac{\alpha}{2}+n\alpha)^2,
\end{equation}
in this parametric domain, both the energy spectra change. 

}
\end{itemize}
  It is worth mentioning that the spectral changes can be viewed as the transformation $n\rightarrow -n$. This is identical to the earlier mentioned symmetry $m\rightarrow -m$, in the P\"oschl-Teller potential. Analogously, $A\rightarrow -A+ \alpha$ is akin to the symmetry $l\rightarrow -(l+1)$.
In the following, we illustrate the relationship between superpotenial and modified KdV equation. In subsequent subsection, we discuss about identifying the correct ground state emerging from above two superpotentails and various regions of $\mathcal{PT}$ phases.

\subsection{Isospectral deformation and connection with modified KdV equation}
The two different superpotentials, giving rise to the same potential, characterise iso-spectral deformation in supersymmetric quantum mechanics. In this, a given potential remains unchanged by an additive deformation of the superpotential, whereas its partner potential is changed having the same energy levels. It is to be noted that superpotential is dependent on the ground state that encodes the boundary conditions, which the potential by itself does not \cite{CAP}. Thus isospectral deformation only keeps the value of potential ($V_{\mathcal{PT}}$) unchanged, not the system boundaries. It is well-known that, iso-spectral deformation of the superpotentials ${\tilde W} (x) = W(x) +\phi (x)$, leads to  the  Bernoulli equation:
\begin{equation}\label{bb}
    \frac{d\phi(x)}{dx}-2W_2(x)\phi(x)-\phi^2(x)=0.
\end{equation}
For the real Scarf II, $\phi(x)$ satisfies the modified KdV equation \cite{khare4, MSP}. 
In the case of complex superpotential potential, $W_1(x)=W_2(x)+\phi(x),$ where 
\begin{equation}
\begin{split}
    \phi(x)&= (A-B+ \alpha/2)~tanh\,\alpha x + i~(B-A-\alpha/2) ~sech\,\alpha x\\
    &\equiv 
    C~tanh\,\alpha x + i~D~sech\,\alpha x,
    \end{split}
\end{equation}
 the modified KdV equation is  satisfied,
\begin{equation}
\frac{ \partial\phi}{\partial t} - \phi^2\, \frac{ \partial\phi}{\partial x} + \frac{ \partial^3\phi}{\partial x^3} = 0,
\end{equation}
provided conditions $C=\pm D$. It is evident that condition $C = -D$  is identically satisfied in the case of this present complex superpotential.

\subsection{Parameter sensitivity and bifurcation phenomena }
	We now consider the energy eigenvalues in Eqs. (\ref{E1}) and (\ref{E2}) defined for the two superpotentials in Eq. (\ref{W1}) and Eq. (\ref{W2}) respectively. The lowest energy in both the cases are $-A^2$ and $-(B-\frac{\alpha}{2})^2$ respectively. Therefore, depending on the parameters $A$ and $B$, we can ascribe the ground state either to $-A^2$ or $-(B-\frac{\alpha}{2})^2$, based on which is lower in energy. Evidently, the superpotential is dependent on the ground state wavefunction, which are different for the two cases and there can be only one ground state in the one dimensional case. Therefore, only the superpotential, corresponding to the lower energy among the mentioned lowest energies, should exist at a given time. The condition for which ground state is allowed with $E_0 = -A^2$,  can be expressed as 
\begin{equation}\label{H1}
\begin{split}
&A^2-(B-\alpha/2)^2=K^2, \quad K = constant;\\
\text{or},\\  &\frac{A^2}{K^2}-\frac{(B-\alpha/2)^2}{K^2}=1,
\end{split}
\end{equation}
which is a hyperbola in the parameter space of $A$ and $B$, with the superpotential given in Eq. (\ref{W1}).
\begin{figure}[h]
\begin{center}
\includegraphics[scale=0.9]{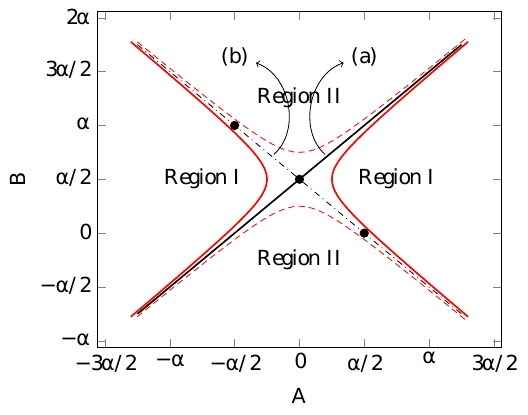}
\caption{Regions I and II correspond to the specific ground states based on the parameters involved. The asymptote with positive slope (a) corresponds to the condition for exceptional point and zero-width resonance while that with negative slope (b) corresponds to isospectral deformation. A small perturbation from the condition for exceptional point or isospectral deformation may lead us to a domain different from the domain of our original trajectory. }
\label{fig21}
\end{center}
\end{figure}
The contrary condition expressed as:
\begin{equation}\label{H2}
\begin{split}
&(B-\alpha/2)^2-A^2=K_1^2, \quad K_1=constant; \\
\text{or},\\ & \frac{(B-\alpha/2)^2}{K_1^2}-\frac{A^2}{K_1^2}=1
\end{split}
\end{equation}
yields a conjugate hyperbola to the previous one, such that $-(B-\frac{\alpha}{2})^2$ is the ground state and the superpotential in Eq. (\ref{W2}) holds. Fig. 2 depicts the allowed parameter domains, corresponding to the EP and the zero-width resonance. The allowed regions are bounded by two hyperbolae, where the asymptotes $B=\pm A+\alpha/2$ point to conditions pertaining to the exceptional point and isospectral deformation respectively. 
The different features of the above figure can be explicated as:
\begin{itemize}
\item{\textbf{Region I} contains all the hyperbolae with $-A^2$ as the ground state with superpotential in Eq. (\ref{W1}).}
\item{\textbf{Region II} contains all the hyperbolae with $-(B-\frac{\alpha}{2})^2$ as the ground state with superpotential in Eq. (\ref{W2}).}
\item{(a) corresponds to the asymptote given by condition $A=B-\alpha/2$ point and the regime of zero-width resonance.}
\item{(b) is the other asymptote but it is given by the condition $A=-B+\alpha/2$ and is a possible region for real energy eigenvalues and it relates the superpotentials in Eq. (\ref{W1}) and Eq. (\ref{W2}) as \textbf{isospectrally deformed} counterparts.}
\item{The coordinate points $(\alpha/2,0)$ and $(-\alpha/2,\alpha)$  are the only two points lying on asymptote $A=-B+\alpha/2$, on which the isospectral deformation satisfy the \textbf{modified KdV equation.}   } 
\end{itemize}
 In region I and region II, each point on $A$ will have two different points on $B$, leading to two  superpontetials indicating a bifurcation behaviour in parameter domain \cite{pkp7}. Similar behaviour is observed for parameter $B$ as well. On the other hand, if for a certain perturbation to the parameters, one reaches the exceptional point then on the opposite path one can move either through the hyperbola one had been originally along or the conjugate hyperbola, making jumps from one superpotential to another possible. As will be explicated in the following section,  two superpotentials which characterize the complex Scarf II potential describe the energy eigenvalues in  regions of the $\mathcal{PT}$-unbroken phase separated by the conditions of isospectral deformation and $\mathcal{PT}$-symmetry breaking, where the two energy eigenvalue sets merge. However, the superpotentials do not merge at the isopectral deformation, while they merge for the $\mathcal{PT}$-symmetry breaking as shown in Fig. 2. The hyperbolic nature of the difference between the two ground state energy representations which asymptotically tend to the $\mathcal{PT}$-breaking show the $\mathcal{PT}$ phase transition.  SUSY broken regions as shown in Fig. 1 can be considered to be arising due to isospectral deformations which follows independently from the Bernoulli equation. Together Fig. \ref{fig20} and Fig. \ref{fig21} depict four parameter domains for energy representation.

 \section{Broken $\mathcal{PT}-$symmetric phase}
For $C^{\mathcal{PT}}\neq 0$, one finds: $A=B-\frac{\alpha}{2}$, which on substituting in Eq. \ref{oo}, yields 
\begin{equation}\label{oos}
W_{\mathcal{PT}}^{\pm}=\Big(A\pm iC^{\mathcal{PT}}\Big)tanh\,\alpha x+\Big(\pm C^{\mathcal{PT}}+i(A+\frac{\alpha}{2})\Big)sech\,\alpha x.
\end{equation}
It is to be noted that both the superpotentials lead to the same $\mathcal{PT}-$symmetric complex potential, corresponding to broken $\mathcal{PT}$-symmetry. Invoking the shape invariance condition, one can  obtain the complex energy spectra,
\begin{equation}
     E_{n}^{\pm}=-(A-n\alpha\pm i C^{\mathcal{PT}})^2.
\end{equation}
The complex conjugate pair of energy eigenvalues are associated with either of the two superpotentials in Eq. \ref{oos}. The corresponding eigenfunctions are given by,
\begin{equation}
\begin{split}
\psi_{n}^{\pm}(x)\propto &\Big(sech\,\alpha x\Big)^{\frac{1}{\alpha}(A\pm iC^{\mathcal{PT}})}\text{exp}\Big[-\frac{i}{\alpha}(A+\frac{\alpha}{2}\mp i C^{\mathcal{PT}})\\
&tan^{-1}(sinh\,\alpha x)\Big]P_{n}^{\mp\frac{2iC^{\mathcal{PT}}}{\alpha},-\frac{2A}{\alpha}-1}[i~sinh\,\alpha x].
\end{split}
\end{equation}
This pair of wavefunctions are related to each other with $\mathcal{PT}$ transformations. Here, $P_{n}^{\mp\frac{2iC^{\mathcal{PT}}}{\alpha},-(\frac{2A}{\alpha}+1)}[i\sinh\,\alpha x]$ is the $n^{\text{th}}$ order Jacobi polynomial. The normalizability of ground state wavefunction requires that SUSY remains intact in broken $\mathcal{PT}$ domain. 
It is also worthwhile to notice that, inspite of being in broken $\mathcal{PT}$-phase,   at points $A=n\alpha$,  complex eigenmodes have energy eigenvalue, $E_{SS}={C^\mathcal{PT}}^2$, constituting the zero-width resonances \cite{AM11} with conditions identical to the one in \cite{ZA12}.

\section{Conclusion}
In conclusion, complex refractive index profile with  $\mathcal {PT}$ and SUSY is investigated for understanding of broken and unbroken phases of the two symmetries and their interconnection. The broken and unbroken phases of $\mathcal {PT}$ have been  found to exist in mutually exclusive domains. For $\mathcal{PT}$ unbroken case, the two superpotentials, interconnected by a parametric transformation, lead to a unique potential. The parameter domains are identified where $\mathcal{PT}-$symmetry is preserved but SUSY breaking happens. In these domains, one needs to exploit a non-trivial shape invariance to obtain the corresponding energy spectra. On the other hand, the broken phase of $\mathcal{PT}$ accommodates complex energy spectra with preserving SUSY in the entire parameter range, yielding zero-width resonance states at integral values of $A$. Moreover, the parametric conditions for isospectral deformation, connecting the two superpotentials describing the same potential in the $\mathcal{PT}$-unbroken phase, is found to be associated with the modified KdV equation of a specific kind. The intersecting lines clearly demarcate the $\mathcal{PT}$-symmetry breaking boundary as well as the isospectral deformation criteria, along which lie points connecting the deformations associated with the modified KdV equation.

Recently $\mathcal{PT}-$symmetric crystals have been experimentally realized with  bulk boundary dynamics and topology, in photonic graphene \cite{graphene}. $\mathcal{PT}-$symmetric phonon lasers have also been proposed theoretically \cite{JOL} and shown experimentally  in optomechanical systems \cite{china}, which will enable $\mathcal{PT}$-symmetry phononics akin to the photonic domain. $\mathcal{PT}$-symmetry has been observed in dissipative systems in ultracold atoms \cite{luo}. All these systems need careful study both from symmetry prospective as well as for technical applications \cite{saxena}. $\mathcal{PT}$ symmetry have manifested in kicked top \cite{graefe1}, intensity-intensity correlation \cite{graefe2} and other physical systems \cite{jog}. The physical phenomena near the exceptional points have been in the forefront area of research, revealing many counter-intuitive effects associated with it. Similarly, dynamical systems with solitonic excitations have been under investigation for the effect of $\mathcal{PT}-$symmetry on this extended objects. We hope to use the understanding gained from this exactly solvable system for throwing light on some of these phenomena in near future.\\ \vskip 0.8cm
\noindent
{\large\bf {Acknowledgments}}\vskip 0.1cm \noindent
AS and PKP acknowledge the support from DST, India   through Grant No.: DST/ICPS/QuST/Theme-1/2019/2020-21/01.

\end{document}